# EFFICIENT ALGORITHMS FOR SEVERAL CONSTRAINED RESOURCE ALLOCATION, MANAGEMENT AND DISCOVERY PROBLEMS

**Mugurel Ionut Andreica**
Politehnica University of Bucharest
mugurel.andreica@cs.pub.ro

**Madalina Ecaterina Andreica**
The Bucharest Academy of Economic Studies
madalina.andreica@gmail.com

**Daniel Ardelean**
Commercial Academy Satu Mare
dgardelean@yahoo.com

**Abstract:** In this paper we present efficient algorithmic solutions for several constrained resource allocation, management and discovery problems. We consider new types of resource allocation models and constraints, and we present new geometric techniques which are useful when the resources are mapped to points into a multidimensional feature space. We also consider a resource discovery problem for which we present a guessing game theoretical model.

## 1. Introduction

Resource allocation and management is a crucial aspect in many domains, like production scheduling, merchandise distribution, business planning, distributed computing, and so on. Many resource allocation, management and discovery problems have been studied in the literature and many models were proposed. In this paper we consider several such problems with various constraints, for which we develop novel, efficient algorithmic solutions. In Section 2 we discuss related work. In Sections 3-6 we present the considered problems together with the proposed solutions, and in Section 7 we conclude.

## 2. Related Work

Many resource allocation, management and discovery models and algorithms were proposed in the literature. Some economic problems concerning the distribution of resources (usually financial - among competing groups of people or programs) are usually described as a multi-attribute or multi-objective decision making problem [4, 8], for which there are several methods and algorithms that lead to the optimal choice in different cases of certainty, risk or uncertainty. In some cases even econometrical tools can help in the allocation process. However, the resource allocation problem was more intensely studied in the operational research fields. For example, a resource allocation problem with just a few resources and activities, can be solved by simply using a Gantt graph, while for more complex problems, several heuristic algorithms were developed. In such cases, usually an ADC-time problem is firstly solved and then all the activities are allocated at the earliest beginning time moment so that no resource exceeds the available limits. Otherwise they are postponed by using some priority rules. Another possible approach of the resource allocation problems can be that of a recursive optimization problem. In some cases, an optimal resource allocation problem can be solved by using the dynamic programming methodology, that consists of the fact that after specifying the objective function that needs to be maximized or minimized, the constraints and an appropriate initial condition for the state of control variable, the problem can be re-described by a recursive relation (also known as the Bellman equation) and then solved [1, 3]. Resource usage optimization problems modeled as single-player games were considered in [2]. Search procedures similar to the ones we present in Section 5 were presented in [5, 6].

## 3. Allocating Resource 3-Tuples

We have $N$ (physical or virtual) resources, each of which having a certain resource amount $x(i)$ ($1 \leq i \leq N$). We want to choose $K$ 3-tuples ($1 \leq K \leq N/3$), such that every resource is part of at most one 3-tuple. Let's assume that we selected a 3-tuple with the amounts of resources $A$, $B$ and $C$ ($A \leq B \leq C$). The cost of the 3-tuple is $|B-A|^P$ ($1 \leq P \leq 10$) ($A$ and $B$ are the *special* values of the 3-tuple). We want to choose the $K$ 3-tuples such that the sum of the costs of the 3-tuples is minimum.

We will first sort all the resources, such that we have $x(1) \leq \ldots \leq x(N)$. A careful analysis leads to the conclusion that the two special values $A$ and $B$ of a 3-tuple must be two consecutive values in the sorted order of the resource amounts (e.g. $x(i)$ and $x(i+1)$). The proof of this fact begins by showing that if the two special values that determine the cost of *2* different 3-tuples ($x(p)$ and $x(q)$, respectively $x(u)$ and $x(v)$) have the property that the intervals $[p,q]$ and $[u,v]$ are not disjoint, then the 3-tuples can be modified such that the intervals are disjoint and the total cost does not increase (e.g. we sort the four values $u, v, p, q$ as $e<f<g<h$ and we replace the two special values in the 3-tuple previously containing $x(h)$ by $(x(g), x(h))$ and the two special values in the other 3-tuple by $(x(e), x(f))$. Then, we can easily notice that the intervals formed by the two special values of each 3-tuple should be as small as possible, i.e. they should consist of two values which are on consecutive positions in the sorted order.

The first solution idea consists of computing $Cmin(i,j)=$ the total cost needed for choosing $j$ 3-tuples from among the first $i$ resources. $Cmin(i \geq 0, 0)=0$ and $Cmin(1,j>0)= +\infty$. $Cmin(i \geq 2, j \geq 1)$ can be computed as $min\{Cmin(i-1,j)$ (if $x(i)$ is not one of the 2 special values of a 3-tuple), $Cmin(i-2,j-1) + |x(i)-x(i-1)|^P$ (if $x(i-1)$ and $x(i)$ are the two special values of a 3-tuple)}. However, we notice that, this way, we do not have the certainty that we will be able to add the third resource to the newly formed 3-tuple (as the third resource should have an amount which is larger than or equal to the two special values). We could enhance the solution by adding an extra parameter $r$ at the dynamic programming state, representing the number of 3-tuples which are still incomplete (i.e. only two special values were added). Thus, $Cmin(i≥1,j≥1,0≤r≤j)= min\{Cmin(i-1, j, r+1), Cmin(i-1, j, r), Cmin(i-2, j-1, r-1)+|x(i)-x(i-1)|^P\}$. However, the time complexity now becomes $O(N·K^2)$.

We can maintain an $O(N·K)$ time complexity as follows. We will traverse the resources from the last towards the first. $Cmin(i,j)$ will now be the minimum total cost of choosing $j$ 3-tuples among the resources $i, i+1, …, N$. $Cmin(*,0)=0$ and $Cmin(N≤i≤N+1, j≥1)=+∞$. For $1≤i≤N-1$ and $1≤j≤K$ we have: if $(N-i+1<3·j)$ then $Cmin(i,j)=+∞$; otherwise, $Cmin(i,j)=min\{Cmin(i+1,j), |x(i+1)-x(i)|^P+Cmin(i+2,j)\}$. The minimum total cost is $Cmin(1,K)$ and the time complexity is $O(N·K)$.

## 4. Resource Collector

We consider a resource collection optimization problem, which we model using terms borrowed from game theory. We have a game which consists of $N$ piles, each of them containing $m_1, m_2, …, m_N$ recipients containing an important resource; the recipients are placed in a stack. Each recipient will be identified by $(i,j)$, where $i$ is the pile number and $j$ is its order in the pile ($j=1$ means that the recipient is at the bottom of the pile). Each recipient $(i,j)$ contains an amount of resources equal to $z_{i,j}·c_{i,j}$ resources. At every move, the player can remove a recipient from the top of a pile and collect all the resources inside that recipient. After each move, the amount of resources inside every remaining recipient $(i,j)$ decreases by $c_{i,j}$. If the amount of resources inside the recipient becomes $0$, the recipient *"magically"* disappears from the pile. The player's objective is to collect the largest possible quantity of resources. The game obviously ends when there are no more recipients in any pile. In order to compute the best strategy, we will use dynamic programming and compute the following values: $C_{max}[t, b_1, b_2, …, b_N]$=the maximum quantity of resources which can be collected from now on if, at the beginning of move $t$ ($t≥1$), the recipient $(i,b_i)$ is at the top of pile $i$ ($0≤b_i≤m_i$, $1≤i≤N$). We will compute these values in descending order of the parameter $t$ and, for each $t$, in ascending lexicographic order of the tuples $(b_1,…,b_N)$. The maximum value for $t$ is $m_1+m_2+…+m_N+1$. The maximum collected quantity of resources is $C_{max}[1, m_1,…,m_N]$. The equation presented below (eq. 1) can be translated into an algorithm with $O((m_1+m_2+…+m_N)·(m_1+1)·…·(m_N+1)·N)$ complexity, which can be used only for a small number of piles and a small number of recipients in each pile.

We can also obtain the optimal strategy (which recipient to choose at each move) from the values $C_{max}$, by tracing the way these values were computed. We can easily see how the game considers actions taken upon any valuable resource whose reserves deplete at constant rates. Furthermore, the resource reserves are arranged in a stack-like fashion (for instance, as layers in the ground) and can only be accessed from the top to the bottom. Particular situations arise if the resource amounts do not decrease from the recipients, in which case all the initial resources will eventually be collected, or if the recipients do not disappear when they contain no more resources. In this second case, $C_{max}[t, b_1, …, b_N]$ depends only on the values $C_{max}[t+1, *, …, *]$ in eq. (1) presented below. In this case we can remove the parameter $t$ from the state definition, because it can be immediately determined from the values $b_1, …, b_N$, as: $t=1+(m_1-b_1)+…+(m_N-b_N)$.

$$C_{max}[t, 0, 0, …, 0]=0$$
$$C_{max}[t,b_1,b_2,...,b_N]=$$

$$\max\begin{cases} \max\begin{cases} C_{max}[t,b_1-1,b_2,...,b_N], \text{if } t>z_{1,b_1} \text{ and } b_1>0 \\ C_{max}[t,b_1,b_2-1,...,b_N], \text{if } t>z_{2,b_2} \text{ and } b_2>0 \\ ... \\ C_{max}[t,b_1,b_2,...,b_N-1], \text{if } t>z_{N,b_N} \text{ and } b_N>0 \end{cases} \\ \max\begin{cases} C_{max}[t+1,b_1-1,b_2,...,b_N]+c_{1,b_1}·(z_{1,b_1}-t+1), \\ \quad \text{if } t≤z_{1,b_1} \text{ and } b_1>0 \\ C_{max}[t+1,b_1,b_2-1,...,b_N]+c_{2,b_2}·(z_{2,b_2}-t+1), \\ \quad \text{if } t≤z_{2,b_2} \text{ and } b_2>0 \\ ... \\ C_{max}[t+1,b_1,b_2,...,b_N-1]+c_{N,b_N}·(z_{N,b_N}-t+1), \\ \quad \text{if } t≤z_{N,b_N} \text{ and } b_N>0 \end{cases} \end{cases} \quad (1)$$

## 5. Inter-Point Distances in the $L_1$ and $L_∞$ Metrics

Resource allocation and management techniques occasionally model the resources as points in a multi-dimensional space (where each dimension corresponds to a property of the resources). In these cases, distance queries are very frequent, when searching for some resources which are close to points corresponding to some specific features. In this section we consider the following multidimensional geometric problems. We have N points in d-dimensional space. Every point $i$ ($1≤i≤N$) has the coordinates $(x(i,1), …, x(i,d))$. The distance between 2 points is considered to be: (1) for the case $d≤2$, $L_1$ or weighted $L_∞$; (2) for $d≥3$, weighted $L_∞$. We are interested in computing efficiently the $K^{th}$ smallest distance between any pair of points ($1≤K≤N·(N-1)/2$). The $L_1$ distance between 2 points $(x_1,y_1)$ and $(x_2,y_2)$ is $|x_1-x_2|+|y_1-y_2|$. The weighted $L_∞$ distance between 2 points $(x(i,1), …, x(i,d))$ and $(x(j,1), …, x(j,d))$ is $max\{w(1)·|x(i,1)-x(j,1)|, …, w(d)·|x(i,d)-x(j,d)|\}$ (for $d$ given weights $w(1), …, w(d)$).

For $d=1$, both $L_1$ and $L_∞$ are equivalent. Let's notice that they are equivalent for $d=2$, too. For the $L_∞$ distance, the points which are at a distance $≤E$ from a point $(x,y)$ are located within a square with side lengths $2·E$, the center at $(x,y)$, and the sides are parallel to the coordinate axes. For $L_1$, the points at a distance $≤E$ from a point $(x,y)$ are located inside a square of side length $E·sqrt(2)$, the center at $(x,y)$, and with its sides rotated by $45$ degrees from the coordinate axes. Thus, if we rotate all the points by $45$ degrees around the origin, 2 points $p_1$ and $p_2$ are at a $L_1$ distance $≤E$, only if the rotated points $p_1'$ and $p_2'$ are at a $L_∞$ distance $≤E·sqrt(2)/2$. In conclusion, we can consider only the $L_∞$ case (if the $K^{th}$ smallest distance in this case has the value $Z$, then the corresponding $L_1$ distance has the value $Z·sqrt(2)$).

We will binary search the value of the $K^{th}$ smallest distance $DK$. Let $D_{cand}$ be the value chosen in the binary search at some step. We will compute $nd(D_{cand})$=the number of pairs of points which are at distance $≤D_{cand}$. If

$nd(D_{cand}) \geq K$, then $D_{cand} \geq DK$; if $nd(D_{cand}) < K$, then $D_{cand} < DK$. In order to compute $nd(D_{cand})$, we will proceed as follows. We will assign to each point a weight $w(i)=1$. Then, we will construct a multidimensional tree (range tree or kd-tree) over the $N$ points. With the help of such a tree, we will be able to compute efficiently the sum of the weights of the points from any multidimensional range (when the weights are $1$, we actually compute the number of points located in that range). Every tree node maintains the sum of the weights of the points in its subtree (for the range tree, this value is maintained only in the trees corresponding to the last dimension; the trees for the other dimensions are only used for indexing, directing the search, and for partitioning the search interval in every dimension). At a search we will compute the sum of the values stored in every visited node whose associated interval (range) is fully contained in the search range. For a range tree, the time complexity of a search is $O(log^d(N))$ (or, when all the weights are $1$, $O(log^{min\{1,d-1\}}(N))$ if we use fractional cascading). We will consider every point $i$, one at a time. We will compute the number of points located in the multidimensional hyper-rectangle with side lengths $2 \cdot D_{cand}/w(j)$ (for every dimension $j=1,...,d$), with the center at $(x(i,1), ..., x(i,d))$ (let this value be $np(i, D_{cand})$). $nd(D_{cand})$ is equal to $(np(1,D_{cand})+...+np(N,D_{cand})-N)/2$. For $d=1$ we only need to sort the points according to the x-coordinates and binary search the smallest index $A$ and the largest index $B$ of a point located at "normal" (unweighted) distance $\leq D_{cand}/w(1)$ from every point $i$: $np(i,D_{cand})=B-A+1$. The time complexity of computing $nd(D_{cand})$ can be reduced, as follows. We will construct a multidimensional tree over the dimensions $1, ..., d-1$ of all the $N$ points, in which every point initially has weight $0$. Then, we will sort the points in ascending order of their $x(*,d)$ coordinates and we will sweep them (from $-\infty$ to $+\infty$) with a $(d-1)$ dimensional zone composed of two parallel hyper-planes, orthogonal to the dimension $j$ and located at (unweighted) distance $2 \cdot D_{cand}/w(d)$ from each other. We assign $3$ events to each point: a *zone entrance* event, a *zone exit* event, and a *zone query* event. If the zone is characterized by a coordinate $xz$ in dimension $d$ (the $2$ hyper-planes are at the coordinates $xz$ and $xz-2 \cdot D_{cand}/w(d)$), the zone entrance event of a point $i$ will be characterized by $(xze=x(i,d), i, IN)$, the zone exit event will be characterized by $(xze=x(i,d)+2 \cdot D_{cand}/w(d), i, OUT)$ and the zone query event will be $(xze=x(i,d)+D_{cand}/w(d), i, QUERY)$. We will sort the events increasingly, according to their $xze$ coordinates (if multiple events have the same $xze$ coordinate, we will place in the sorted order the entrance events, then the query events and then the exit events at the same $xze$ coordinate). We will then traverse the events in the sorted order. When we reach an event $(xze, i, IN)$, we will set the weight of point $i$ from the multidimensional tree to $1$ (remember that only the first $d-1$ coordinates of point $i$ were considered in the multidimensional tree); this will take $O(log^{d-1}(N))$ time. When we reach an event $(xze, i, QUERY)$, we consider a multidimensional hyper-rectangle with side lengths $2 \cdot D_{cand}/w(j)$ (for every dimension $j$; $1 \leq j \leq d-1$), with the center at $(x(i,1), ..., x(i,d-1))$, and we will compute the sum of the weights of the points in the multidimensional tree which are located in this hyper-rectangle (using the previously described technique, in $O(log^{d-1}(N))$ time for a range). For a zone exit event $(xze, i, OUT)$, we will set the weight of the point $(x(i,1), ..., x(i,d-1))$ from the multidimensional tree back to $0$ (in $O(log^{d-1}(N))$ time). The presented techniques can be used in order to solve other problems, too. For instance, if we want to know how many pairs of points have distances whose values belong to the interval $[A,B]$, we compute $U=nd(A-\varepsilon)$ and $V=nd(B)$; the result will be $V-U$ ($\varepsilon>0$ is a small constant).

The case $K=N \cdot (N-1)/2$ can be solved easily for $L_1$ and $L_\infty$ for any number of dimensions $d$. For $L_\infty$ we consider every dimension $j$ separately and we compute $xmax(j)=max\{x(i,j)|1 \leq i \leq N\}$ and $xmin(j)=min\{x(i,j)|1 \leq i \leq N\}$. The largest distance is $max\{w(j) \cdot (xmax(j)-xmin(j))|1 \leq j \leq d\}$. The time complexity of the algorithm is $O(N \cdot d)$. The weighted distance $L_1$ between $2$ points $(x(i,1), ..., x(i,d))$ and $(x(j,1), ..., x(j,d))$ is $w(1) \cdot |x(i,1)-x(j,1)|+...+w(d) \cdot |x(i,d)-x(j,d)|$. We will transform every point $i$ into a point $p(i)$ which has $2^d$ coordinates (using a technique presented in [7]). In order to compute the coordinates of $p(i)$ we will consider all the tuples $(s(1), ..., s(d))$ (with $s(j)=-1$ or $+1$; $1 \leq j \leq d$) in ascending lexicographic order. For the $j^{th}$ such tuple we obtain the $j^{th}$ coordinate $p(i,j)=w(1) \cdot s(1) \cdot x(i,1)+...+w(d) \cdot s(d) \cdot x(i,d)$. The unweighted $L_\infty$ distance between the $p(i)$ points (computed according to the algorithm described above) is equal to the maximum weighted $L_1$ distance of the initial points. Thus, we obtained an algorithm with $O(N \cdot d \cdot 2^d + N \cdot 2^d)$ time complexity. If we could compute the $p(i,j)$ values more efficiently, we could reach an $O(N \cdot 2^d)$ time complexity. We will use a recursive function for generating the tuples $(s(1), ..., s(d))$, maintaining the partial sum $SP$ of the current prefix of the tuple. We will also maintain a counter $T$, for counting the number of generated tuples $(s(1),...,s(d))$. We call $GenTuples(i,1)$ for every point $i$ ($1 \leq i \leq N$) (after setting $SP=T=0$ before every call).

**GenTuples(i, j):**
**if** $(j=d+1)$ **then** $\{ T=T+1; p(i,T)=SP \}$
**else** $\{ s(j)=-1; SP=SP+w(j) \cdot s(j) \cdot x(i,j); GenTuples(i, j+1);$
    $SP=SP-w(j) \cdot s(j) \cdot x(i,j);$
    $s(j)=+1; SP=SP+w(j) \cdot s(j) \cdot x(i,j); GenTuples(i, j+1);$
    $SP=SP-w(j) \cdot s(j) \cdot x(i,j); \}$

A related problem is the following. Given $N$ points in d-dimensional space, find the largest value of a factor $F$, with the property that, if we built $N$ hyper-rectangles with side lengths $w(j) \cdot F$ (in every dimension $j$) and with the centers in the $N$ given points, these hyper-rectangles would not intersect (at most, they would only touch each other). We will binary search the optimal value $F_{opt}$. Let $F_{cand}$ be the value selected during a step of the binary search. We will consider a grid with unit lengths equal to $L(j)=w(j) \cdot F_{cand}$ (in dimension $j$; $1 \leq j \leq d$). We compute a tuple $z(i)=(z(i,j)=floor(x(i,j)/L(j))$ ($1 \leq j \leq d$) for every point $i$. We will maintain a data structure (e.g. a hash table or a balanced binary search tree) where we will insert the pairs $(key=z(i), value=(x(i,1), ..., x(i,d)))$. If we introduce two pairs with identical keys, then there will be $2$ points in the same *grid cell* and, thus, the two hyper-rectangles corresponding to the two keys have a non-null intersection. Thus, $F_{cand}$ would not be a feasible value. If we didn't find $2$ identical tuples $z(i)$, then we will consider every point $i$ and we will verify, for each of the $3^d-1$ tuples $(s(1), ..., s(d))$ (with $s(j)=-1, 0$ or $1$; $1 \leq j \leq d$; but not all the $s(j)$ values may be $0$), if the tuples

$z'(i)=(z(i,1)+s(1), ..., z(i,d)+s(d))$ were inserted into the data structure. For each verified tuple $z'(i)$ which exists in the data structure we obtain (from the data structure) the point $(x(j,1), ..., x(j,d))$ for which the tuple $z'(i)$ is the key. We now verify if the hyper-rectangles associated to the points $(x(i,1), ..., x(i,d))$ and $(x(j,1), ..., x(j,d))$ intersect each other; if they do, then $F_{cand}$ is not feasible. If no intersection is found, then $F_{cand}$ is feasible. If we decided that $F_{cand}$ is not feasible, then we will consider smaller values in the binary search (because $F_{cand}>F_{opt}$); otherwise, we will consider larger values (because $F_{cand} \leq F_{opt}$). The time complexity of this algorithm is $O(N \cdot 3^d)$ (if we use hash tables) or $O(N \cdot 3^d \cdot log(N))$ (if we use balanced binary search trees, or if we simply sort the tuples $z(i)$ and binary search every tuple $z'(i)$).

## 6. Guessing a Permutation

We consider the following resource discovery problem, modeled as guessing game. We have $n$ resources, each of which has one value between $1$ and $n$ and their values are distinct (i.e. the values form a permutation of $\{1,...,n\}$). A player has to find the secret permutation $S$, by asking questions of the form $Ask(p)$, where $p$ is a permutation with $n$ elements. The answer is an array $ans$ with $n$ elements, where $ans(i)=0$ if $p(i)=S(i)$, $ans(i)=-1$ if $p(i)<S(i)$, and $ans(i)=+1$ if $p(i)>S(i)$ ($p(i)$ and $S(i)$ denote the $i^{th}$ element of the permutations $p$ and $S$, respectively). We want to find $S$ using a strategy which minimizes the total number of questions in the worst-case.

We will use the uncertainty minimization principle, which we introduce next. We assign an uncertainty value $U(S)$ to the current state $S$ of the game. The state $S$ is based on the answers received to the previously asked questions. At each step, we consider every type of question $Q$ that we may ask. For each question, we consider all the possible answers $A$ to this question and we evaluate the state $S'$ of the game in case we ask the question $Q$ and receive the answer $A$, and its uncertainty value $U(S')$. The weight of the question $Q$, $w(S,Q)$ is $max\{U(S') \mid S'$ is a state which is reached by asking question $Q$ in state $S$ and receiving one of its possible answers, which is consistent with the previous answers$\}$. In state $S$, we will ask the question $Q$ with the minimum value of $w(S,Q)$ (i.e. that question $Q$ for which the worst case uncertainty is as small as possible). The game ends when we reach a state $S$ with $U(S)=0$, i.e. there is no uncertainty regarding the answer that we seek.

We will assign to each position $i$ ($1 \leq i \leq n$) of the permutation an interval $[a(i), b(i)]$, representing the set of values to which $S(i)$ may be equal. Initially, $a(i)=1$ and $b(i)=n$ for every position $i$ ($1 \leq i \leq n$). The uncertainty of a position $i$ is $UP(i)=(b(i)-a(i))$. The uncertainty of a state of the game is equal to the sum of the values $UP(i)$ ($1 \leq i \leq n$). When choosing the next question $Ask(p)$ to ask, we evaluate what would be the new uncertainty $UP(i)$ of every position $i$, in the worst-case, if $p(i)=j$ (for every $1 \leq i,j \leq n$). We denote by $UP_{new}(i,j)$ the new uncertainty of position $i$, if $p(i)=j$. If $j<a(i)$ or $j>b(i)$ or $a(i)=b(i)$, then $UP_{new}(i,j)=b(i)-a(i)$. If $a(i)<b(i)$ then $UP_{new}(i, a(i))=b(i)-a(i)-1$ (in the worst case, the answer would indicate that $p(i)>a(i)$) and $UP_{new}(i,b(i))=b(i)-a(i)-1$ (in the worst case, the answer would indicate that $p(i)<b(i)$). For $a(i)<j<b(i)$, we have $UP_{new}(i,j)=max\{0, b(i)-(j+1), j-1-a(i)\}$ (the three values correspond to the cases $ans(i)=0$, $ans(i)=-1$ and $ans(i)=+1$). We now need to find a permutation $p$ such that the sum of the values $UP_{new}(i,p(i))$ is minimum. Afterwards, we will ask the question $Ask(p)$. After receiving the answer, we modify the values $a(i)$ and $b(i)$ accordingly. If $p(i)=S(i)$, then $a(i)=b(i)=p(i)$; if $p(i)<S(i)$, then $a(i)=p(i)+1$; if $p(i)>S(i)$, then $b(i)=p(i)-1$.

In order to find the optimal permutation $p$, we construct a bipartite graph with $n$ vertices on both sides ($l(1), ..., l(n)$ are the left side vertices and $r(1), ..., r(n)$ are the right side vertices). We add an edge between every pair of vertices $(l(i), r(j))$ and assign to it the cost $UP_{new}(i,j)$. We will now compute a minimum cost maximum matching in this bipartite graph (which can be achieved in $O(n^3)$ time). Then, if $l(i)$ is matched to $r(j)$, we will have $p(i)=j$.

## 7. Conclusions and Future Work

In this paper we considered several constrained resource management, allocation and discovery problems, for which we introduced novel, efficient algorithmic techniques which solve the problems optimally or nearly optimally. The problems were motivated by real-life scenarios, but were tackled mostly from a theoretical perspective. As future work, we will focus on solving problems with immediate applications in practical settings (e.g. by extending some of the solutions which we proposed in this paper).